\def\year{2022}\relax
\documentclass[letterpaper]{article} % DO NOT CHANGE THIS
\usepackage{aaai22}  % DO NOT CHANGE THIS
\usepackage{times}  % DO NOT CHANGE THIS
\usepackage{helvet}  % DO NOT CHANGE THIS
\usepackage{courier}  % DO NOT CHANGE THIS
\usepackage[hyphens]{url}  % DO NOT CHANGE THIS
\usepackage{graphicx} % DO NOT CHANGE THIS
\usepackage{natbib}  % DO NOT CHANGE THIS AND DO NOT ADD ANY OPTIONS TO IT
\usepackage{caption} % DO NOT CHANGE THIS AND DO NOT ADD ANY OPTIONS TO IT
\DeclareCaptionStyle{ruled}{labelfont=normalfont,labelsep=colon,strut=off} % DO NOT CHANGE THIS
\usepackage{amsmath}
\title{Prediction of $\textrm{CO}_2$ Adsorption in Nano-Pores with Graph Neural Networks}
\author {
    Guojing Cong\thanks{Notice: This manuscript has been authored in part by
  UT-Battelle, LLC under Contract No. DE-AC05-00OR22725 with the
  U.S. Department of Energy. The United States Government retains and
  the publisher, by accepting the article for publication,
  acknowledges that the United States Government retains a
  non-exclusive, paid-up, irrevocable, world-wide license to publish
  or reproduce the published form of this manuscript, or allow others
  to do so, for United States Government purposes. The Department of
  Energy will provide public access to these results of federally
  sponsored research in accordance with the DOE Public Access Plan
  (http://energy.gov/downloads/doe-public-access-plan).},\textsuperscript{\rm 1}
    Anshul Gupta,\textsuperscript{\rm 2}
    Rodrigo Neumann,\textsuperscript{\rm 3}
    Maira de Bayser,\textsuperscript{\rm 3}
    Mathias Steiner,\textsuperscript{\rm 3}
    Breannd\'an \'O Conch\'uir \textsuperscript{\rm 4,5}
}
\affiliations {
    \textsuperscript{\rm 1} Oak Ridge National Lab, 1 Bethel Valley Road, Oak Ridge, Tennessee, 37830, U.S.A.\\
    \textsuperscript{\rm 2} IBM Research, T.J. Watson Research Center, 1101 Kitchawan Road, Yorktown Heights, New York, 10598, U.S.A.\\
    \textsuperscript{\rm 3} IBM Research, Av. Rep\'ublica do Chile, 330, Rio de Janeiro, 20031-170, RJ, Brazil\\
    \textsuperscript{\rm 4} IBM Research Europe, The Hartree Centre, Daresbury Laboratory, Warrington, WA4 4AD, U.K.\\
    \textsuperscript{\rm 5} School of Chemical Engineering and Analytical Science, University of Manchester, Manchester M13 9PL, U.K.\\
    congg@ornl.gov, anshul@us.ibm.com, \{rneumann, mgdebayser, mathiast\}@br.ibm.com, breanndan.conchuir@ibm.com
\vspace{-0.75in}
}

\begin{document}

\maketitle

\begin{abstract}
We investigate the graph-based convolutional neural network approach for predicting and ranking gas adsorption properties of crystalline Metal-Organic Framework (MOF) adsorbents for application in post-combustion capture of $\textrm{CO}_2$. Our model is based solely on standard structural input files containing atomistic descriptions of the adsorbent material candidates. We construct novel methodological extensions to match the prediction accuracy of classical machine learning models that were built with hundreds of features at much higher computational cost. Our approach can be more broadly applied to optimize gas capture processes at industrial scale.
\end{abstract}

\section{Introduction}

The discovery of new materials for carbon capture is a major research challenge for its technical complexity and potential global impact. There are potentially millions of materials that are candidates for adsorbents of $\textrm{CO}_2$ from the flue gas of point-source emitters, such as fossil fuel-based power plants. Identifying the most suitable material for a given post-combustion capture scenario requires screening of an inordinate number of material candidates. Among the classes of solid-state $\textrm{CO}_2$ adsorbents, Metal-Organic Frameworks (MOFs) stand out for their chemical diversity and customisable porous structure~\cite{moosavi2020understanding}. 
While experimental screening is rendered unviable by its high cost, computational screening that uses physics-based simulations presents significant challenges. Since it is not feasible to perform molecular-level simulations for millions of material candidates, novel hierarchical screening methods for scientific inference in materials research are required. Existing materials screening approaches contain a top layer in which rapid geometric and topological characterisations of the materials are deployed to eliminate samples with less favorable properties. By using this approach, only the properties of the  most promising  material candidates are subsequently calculated using molecular-level physics simulation, significantly reducing discovery time and computational cost. This approach is promising; however, these top layer topological and geometric descriptors can only classify  samples  for  elimination  or  further  study.  Furthermore, they neglect  the  intricate chemical  interactions  between  various  atomic  species  present in the nanopore framework and gas phase, thus limiting the effectiveness of such descriptors as a screening tool. A data-driven framework is, therefore, needed which explicitly accounts for the complete set of geometric, topological, and chemical mechanisms that determine the results of molecular property simulations.

Previous studies have employed a variety of geometric, topological and chemical descriptors to analyse the $\textrm{CO}_2$ adsorption performance of nanoporous materials. Geometric features can include the accessible pore surface area and volume, pore diameter metrics and crystal density~\cite{boyd2016generalized, coudert2016computational, lee2018high, dureckova2019robust, zhang2019machine, deeg2020silico, jablonka2020big, krishnapriyan2020topological}. Topological representations of the materials can be generated using concepts extracted from persistent homology, and clustering of high performance materials with similar topologies can be performed~\cite{lee2017quantifying, lee2018high, zhang2019machine, chung2019persistence, krishnapriyan2020topological}. These topological methods can also be combined with chemical information to enhance correlations between the descriptors and the material adsorption properties~\cite{townsend2020representation}. Recently, in a comprehensive article~\cite{moosavi2020understanding}, a set of 165 geometric and chemical material descriptors were built and combined with classical machine learning techniques to train algorithms for predicting the adsorption performance of samples from several large MOF databases. 

An alternative approach, which frees the material scientist from having to calculate a multitude of descriptors beforehand, is using deep learning methods such as the Crystal Graph Convolutional Neural Network (CGCNN)~\cite{xie2018crystal}. Originally developed for predicting electronic properties of crystalline materials, such as formation energy, absolute energy and band gap, this model learns material properties from the atoms in the crystal and their nearest-neighbour connections. In a derived work~\cite{rosen2021machine}, the original CGCNN formulation was shown to outperform the accuracy of several other descriptor-based regression models in a high-throughput material screening study of electronic properties. Other deep learning models such as an improved CGCNN (iCGCNN) model~\cite{park2020developing}, and a geometric-information-enhanced crystal graph neural network (GeoCGNN)~\cite{cheng2021geometric} have further improved prediction accuracy. 

In one of the few attempts to merge the use of CGCNN with gas adsorption prediction~\cite{wang2020accelerating}, the authors applied high-throughput Grand Canonical Monte Carlo (GCMC) calculations to 10,000 entries in the CoRE-2019 materials database~\cite{chung2019advances} to train a machine learning classifier that distinguishes materials into high- or low-performance categories. The GCMC simulation is a physics-based statistical method that searches for the equilibrium state of a thermodynamic system based on probabilities that depend on the inter-atomic energy between the framework and the adsorbate atoms. This was applied as a pre-screening step for selecting the most promising candidates for further analysis out of a much larger hypothetical MOF database~\cite{wilmer2012large} containing more than 320,000 materials. The CGCNN classifier was able to identify a small fraction of the database, less than $9\%$, as deserving of further investigation via GCMC simulations.

To date, the use of CGCNNs for creating deep learning regression models capable of predicting and ranking the gas adsorption capacity of nano-porous material candidates based on their atomistic description has not been reported. In the following, we demonstrate an enhanced CGCNN model with significantly improved overall accuracy. This surrogate model can replace the Physics-based simulation and compute material properties in a fraction of the time and with a fraction of computational resources.

\section{Enhanced Crystal Graph Convolutions}
\label{sec-conv}

We introduce our Enhanced Crystal Graph Convolutional Neural Network (ECGCNN) in relation to the popular CGCNN~\cite{xie2018crystal}, and describe our neural network architecture augmentations that significantly improve the prediction of $\textrm{CO}_2$ adsorption properties.

In order to build a predictive model based on graph convolutions, the crystal structure of the MOF is represented by a graph that encodes both atomic information and bonding interactions between atoms. The vertices of the graph represent atoms, and the edges represent the atoms' nearest neighbors, determined by the Euclidean distance between the atoms in 3D Cartesian space; i.e., atoms closer than a threshold distance are considered neighbors and will have an edge connecting them. In fact, there are multiple edges between the same pair of neighbors to represent periodicity in the crystal. Feature vectors corresponding to vertices and edges encode atom and bond properties, respectively. A convolutional neural network is built atop the graph to automatically extract representations that are optimum for predicting target physical properties. The reader is referred to the original paper~\cite{xie2018crystal} for a detailed description of CGCNN. 

\subsection{Notation}

$G=(V,E)$ denotes an undirected multigraph defined by a vertex set $V = \{1,2,\ldots,n\}$ representing the atoms of a crystal with an $n$-atom unit cell and edges $(i,j)_k$ representing connections between atoms $i$ and $j$, where $i,j \in V$. $G$ allows multiple edges between the same pair of vertices to represent periodicity of the crystal structure; therefore, $k$ represents the $k$-th bond between $i$ and $j$. A feature vector $v_{i}$ is associated with each vertex $i$ and a feature vector  $e_{i,j,k}$ is associated with each edge $(i,j)_k$.

\subsection{Graph Convolution}

Graph Neural Networks (GCNs) apply convolutions to features according to the relationship defined by the graph. The vanilla GCN by~\citeauthor{kipf2016semi} is a localized first-order approximation of spectral graph convolutions~\cite{kipf2016semi}.
\begin{equation}
v_i^{l+1} = \sigma \left(b^l + \sum_{v_j \in N(i)} C_{v_i,v_j} v_j^l W^l\right ).
\label{eq-vanilla}
\end{equation}

Here $v_i^l$ is the feature vector at vertex $i$ at convolution level $l$. For each vertex $i$, its hidden feature vector at next level $v_i^{l+1}$ is computed by aggregating hidden states of its neighbors $j \in N(i)$ and possibly of the incident edges as well. In Equation~\ref{eq-vanilla}, $C_{i,j}$ is a constant, $b^l$ is the bias vector, and $W^l$ is the weight matrix; the latter two are parameters to be learned. A prominent feature of GCN that is shared by its later extensions is the message passing mechanism that propagates information through the connections among nodes in the graph.

Graph convolutions for crystal structures have a few major differences from typical GCNs. There can be multiple edges between a pair of nodes, and also the incident edges for a node may change depending on how large a neighborhood is considered according to the strength of the bonds.  
 
Existing GCNs for crystal structures emphasizes the importance of edge features in learning. At level $l+1$, the feature vector $v_i^{l+1}$ for vertex $i$ is computed by a vertex convolution function $\mathcal{C}_V$ using the previous level's features from $i$'s neighbors and from all edges incident on $i$ as follows:
\begin{equation}
\label{eqn-cgcnn}
v^{l+1}_i = \mathcal{C}_V\left( v_i^l, v_j^l, \{e_{i,j,k}\} \right).
\end{equation}

Typically $\mathcal{C}_V$ is a linear combination of all features involved with some nonlinear activation function. As long as it aggregates the features from adjacent vertices and edges, $\mathcal{C}_V$'s definition can be chosen to suitably meet the requirements and the characteristics of the application at hand. In the original CGCNN paper, \citeauthor{xie2018crystal} used the following definition: 
\begin{eqnarray}
\label{eqn-conv1}
&\mathcal{C}_V\left( v_i^l, v_j^l, \{e_{i,j,k}\} \right) = v_i^l \nonumber \\
&+ \Sigma_{j,k} \sigma(z^l_{i,j,k} W_c^t + b_c^l) \bigodot g(z_{i,j,k}^l W_s^l +b_s^l),
\end{eqnarray}
where $\bigodot$ denotes element-wise multiplication, $\sigma$ is the Sigmoid function, $g$ is a nonlinear activation function, and $z_{i,j,k}$ is the concatenation of $v_i$, $v_j$, and $e_{i,j,k}$. $W_c$ and $b_c$ are the shared convolution weight matrix and bias vector, and $W_s$ and $b_s$ are the self weight matrix and bias vector, respectively.

Note the subtle differences between Equation~\ref{eq-vanilla} and Equations~\ref{eqn-cgcnn} and~\ref{eqn-conv1}. Equations~\ref{eqn-cgcnn} and~\ref{eqn-conv1} compute $v_i^l$ for a vertex $i$ by first deriving $|\{e_{i,j,k}\}|$ features using both edge features and node features, and then combine them. 

\subsection{Edge Convolution}

Edge features are obviously critical to learning the properties of a crystal structure. Notice that Equation~\ref{eqn-conv1} for CGCNN uses the edge features as input to the convolution, but these features themselves remain unchanged through the convolution steps. As shown in the Results section, introducing parameters (weights) associated with the edges and evolving edge features through an edge-convolution function $\mathcal{C}_E$ yielded a significant improvement in prediction accuracy.

We introduce convolution to the edges as follows: 
\begin{equation*}
e_{i,j,k}^{l+1} = \mathcal{C}_E\left( e_{i,j,k}^l, v_i^l, v_j^l \right),
\end{equation*}
where
\begin{equation*}
\mathcal{C}_E\left( e_{i,j,k}^l, v_i^l, v_j^l \right) = \textrm{activation} \left (e_{i,j,k}^l\bigoplus v_i^l \bigoplus v_j^l \right).
\end{equation*}

Now a convolution layer produces two kinds of outputs, one feature for each vertex, and one feature for each edge. 

\subsection{Attention}

Attention mechanism has been shown to be a powerful mechanism in deep learning, especially in learning with sequences. In deep neural networks, attention mimics cognitive attention and can capture spacial or temporal interdependence between the most relevant parts of the input. In the graph neural networks, an attention function is learned simultaneously with the features~\cite{Velickovic2018GraphAN}. 

An elementary form of attention (or more precisely, a gated mechanism) is implemented in CGCNN shown in Equation~\ref{eqn-conv1} by the Sigmoid function. Note that the added $v_i^l$ is similar to the concept of pass-through connection first introduced in \texttt{ResNet}~\cite{he2016deep}. 
We introduce formally the attention mechanism as follows:
\begin{equation*}
  v_i^{l+1} = \sum_{j,k} \textbf{Softmax}_{j,k} \{z^l_{i,j,k} W_c^l + b_c^l\}  \bigodot g(z_{i,j,k}^l W_s^l + b_s^l)  
\end{equation*}

Here in producing a feature vector $v_i$, the neural network attends to features of all neighbors of $i$. Since the neighbors' features are computed using edge features as well, this mechanism also implicitly attends to the edge features.

\subsection{Multi-head Attention and beyond}

Multi-head attention has been shown to improve the performance of graph convolution~\cite{Velickovic2018GraphAN}. With multi-head attention,  multiple attention mechanisms concurrently attend to the neighbors and incident edges of a vertex. The equation below illustrates one way of implementing multi-head attention (with $M$ heads) to a graph convolution where the outputs of the attention heads are averaged.
\begin{equation*}
v_i^{l+1} = \frac{1}{M} \sum_{m} \sum_{j,k} S^{l,m} \bigodot g(z_{i,j,k}^l W_s^{l,m} + b_s^{l,m}), 
\end{equation*}
where $S^{l,m}=\textbf{Softmax}_{j,k} \{z^l_{i,j,k} W_c^{l,m} + b_c^{l,m}\}$ and $m \in \{1,2,\ldots,M\}$.

Instead of averaging, we leverage another layer of attention mechanism in combining the results from multiple attend heads. We generate two features, $o_i^{l,m}$ and $a_i^{l,m}$, from each attention head $m$ at level $l$ and use them as below:  
\begin{equation*}
v_i^{l+1} = \sum_m \textbf{Softmax}_m \{ a_i^{l,m} \bigodot o_i^{l,m} \}.
\end{equation*}

\subsection{Other Considerations}

Vanilla graph convolution suffers from over-smoothing when the networks become deep. Recently proposed architectures address this deficiency with various techniques~\cite{Chen2020gcnii,Li2021deepgcn}. For example, initial residual and identity mapping are introduced to GCN to overcome over-smoothing, and achieve good performance on standard benchmarks~\cite{Chen2020gcnii}. However, in our experimentation, we observed that such techniques did not bring tangible improvement in predicting $\textrm{CO}_2$ adsorption. This could be in part due to the highly regular structure of crystal graph and the importance of the edge features.     

Some prior work in using graph convolutions to predict the material properties of crystal structures explores incorporating various physical atomic interaction expressions~\cite{chen2019graph,unke2019physnet,cheng2021geometric}, including electronic state wave functions, directly into the neural network.
Unfortunately, such techniques cannot be readily applied to our use case as all relevant pairwise atomic interactions in the adsorption simulation are between individual $\textrm{CO}_2$ gas molecules and individual framework atoms. While the latter are represented by nodes in our network, the former are not explicitly represented and thus their relative proximity and resultant attractive/repulsive interaction cannot be computed. 

\section{Results}
\label{sec-results}

We report the performance of ECGCNN for predicting $\textrm{CO}_2$ gas uptake. 

\subsection{Training Data}

The training dataset was built from existing Crystallographic Information File (CIF) databases~\cite{moosavi2020understanding}. The CoRE-19 database~\cite{chung2019advances} contains 9,525 MOF materials, while the BW-20K database is a subset of the original BW-DB database~\cite{boyd2016generalized} containing only 19,379 out of 320,000 MOFs.

Each CIF file describes the associated material at an atomistic level, including the atomic number ($Z$), atomic coordinates ($x, y, z$), and atomic partial charges ($q$) of each atom in the crystal. It also contains the geometric features of the crystallographic cell, such as the cell lengths ($a, b, c$) and cell angles ($\alpha, \beta, \gamma$). With the atomic coordinates and the periodicity of the crystal cell, one can identify pairs of nearest neighbour atoms and the inter-atomic distance between them. 
The information about cell geometry was not used in the original CGCNN~\cite{xie2018crystal}; including it in training resulted in a slight improvement of our model.

The target adsorption properties of the materials were calculated using Physics-based GCMC simulations that only take the CIF as input. The $\textrm{CO}_2$ gas uptake for each material was calculated at room temperature for high (16 bar) and low (0.15 bar) pressures, simulating adsorption and desorption conditions, respectively. For more details on the GCMC simulations, please refer to~\citeauthor{moosavi2020understanding}~(\citeyear{moosavi2020understanding}).

\subsection{Hyper-Parameters}

In our experiments, we use four layers of graph convolution. The original input vertex and edge feature vectors are the same ones that are employed in~\cite{xie2018crystal}. The vertex features encode several atomic properties and the edge features encode atomic distance. These vectors are first embedded into $64$ dimensional vectors before graph convolution. All hidden features for the vertices and edges are $64$ and $42$ dimensional, respectively. We use three attention heads and \texttt{ReLU} for activation. Average pooling is used to pool from the atom features into a feature for the corresponding crystal. A two-layer multilayer perceptron (MLP) is used to predict the adsorption of the crystal. These fully connected layers are $64 \times 128$, and $128 \times 1$, respectively. In constructing the graph, we regard the neighbors of each atom to be the nearest $12$ atoms within a radius of $8$ \AA. The Adam optimizer used with $10^{-3}$ learning rate and $5 \times 10^{-4}$ weight decay. We train for $50$ epochs with the training set, and report the test accuracy on the test set with the model that achieved the best validation accuracy on the validation set. We randomly split the datasets into training, validation, and test sets. The reported results are averaged over five splits. 

\begin{figure}[t]
  \centering
    \includegraphics[width=0.5\textwidth]{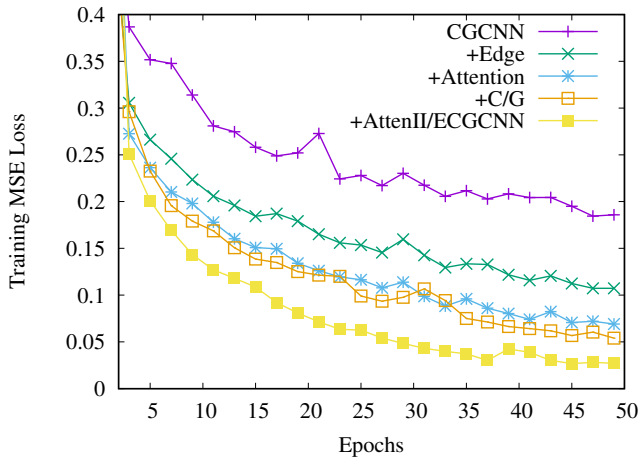}
    \caption{A plot illustrating the corresponding reduction of training MSE loss with the addition of each enhancement to the neural network. These include the original CGCNN, edge convolution, attention, atomic partial charge and geometric information (lengths and angles), and multi-head attention. %The training losses at the end of 50 epochs are recorded in Table~\ref{tab:BW20K}.
    }
    \label{fig:loss-epochs}
\end{figure}

\subsection{Mean Absolute Error and Spearman Rank Correlation Coefficient}

\begin{figure}[t]
  \centering
    \includegraphics[width=0.5\textwidth]{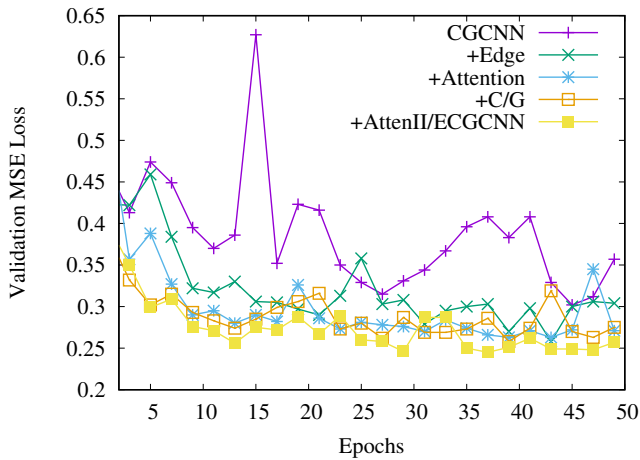}
    \caption{A plot illustrating the corresponding reduction of validation MSE loss with the addition of each enhancement encoded in the neural network.}
    \label{fig:val-loss-epochs}
\end{figure}

The infusion of our enhancements into the CGCNN significantly improves the training performance of the network. Figure~\ref{fig:loss-epochs} shows the evolution of training mean squared error (MSE) loss with the number of training epochs for the various GCNs for the BW-20K dataset at pressure 0.15 bar. In our experiment we start with CGCNN, and then introduce edge convolution, attention mechanism, atom charge information and geometric features, and then multi-head attention incrementally to the plain convolution used in CGCNN. It is clear from the plot that as the neural network becomes more complex and/or more new physics features are included, better training results (lower losses) are achieved. Figure~\ref{fig:val-loss-epochs} shows the evolution of validation MSE loss with the number of training epochs. 

\begin{table*}[bt]
 \begin{center}
 \begin{tabular}{| c | c | c | c | c | c |}
 \hline
 Metric & CGCNN & +Edge & +Attention& +C/G & + AttenII/ECGCNN\\ \hline
 train MSE &  0.197&  0.110& 0.066& 0.058 & 0.023\\ 
 train MAE &  0.304&  0.231&  0.184& 0.175&  0.112\\
 test MAE  & 0.304&  0.260&  0.256& 0.256&  0.239\\
 test SRCC &0.923& 0.931&   0.932& 0.932& 0.943\\ \hline
 \end{tabular} 
\end{center}
\caption{Table assessing the training and test performance of various crystal graph convolutional neural networks on the BW20K 0.15 bar data set~\cite{moosavi2020understanding} after 50 epochs. An 80-10-10 train-validation-test split is deployed throughout. %Note that C/G represents the addition of atomic partial charges and geometric information while Attention II denotes the final multi-head attention enhancement.
}
\label{tab:BW20K}
\end{table*}

The performance of the original CGCNN on this data set, and of the network with each incremental additional enhancement, is detailed in Table~\ref{tab:BW20K}. The networks show a monotonic increase in performance in the training phase, as dictated by an overall reduction of 88\% in the training MSE and 63\% in training mean average error (MAE). This trend is replicated with a reductions of 21\% in the test MAE and a 2\% rise in the test Spearman ranking (SRCC). Note that we deployed a train-test-validation split of 80:10:10 to produce these results recorded in the first five columns of Table~\ref{tab:BW20K}. 

In Table~\ref{tab:CoRE19}, we contrast the test performance of our most enhanced crystal graph convolutional neural network (ECGCNN) to the previous best results obtained for this data set using the application of classical machine learning techniques to geometric and chemical material properties~\cite{moosavi2020understanding}, for all combinations of temperatures and pressures available in this reference. The two middle columns employ the approximate 70-30 test-train split used by~\citeauthor{moosavi2020understanding} while the rightmost column corresponds to the 80-10-10 test-train-validation split used earlier in Table~\ref{tab:BW20K}. Across the four distinct data sets corresponding to two sets of materials simulated under low (0.15 Bar) and high (16 Bar) pressure conditions, one can observe similar performance between our ECGCNN model and the framework outlined by~\citeauthor{moosavi2020understanding} for the first three data sets. This observation holds for both absolute (test MAE) and relative (SRCC) metrics. Both of these performance metrics also improve significantly when one deploys the 80-10-10 test-train-validation split compared to the 70-30 test-train split used in the referenced study. 

\begin{table}[bt]
 \begin{center}
 \begin{tabular}{| c | c | c | c |}
 \hline
 Metric & Moosavi$_{7:3}$ &ECGCNN$_{7:3}$ & ECGCNN$_{8:1:1}$\\ \hline
 \multicolumn{4}{|c|}{BW-20K, 0.15 bar}\\ \hline
 MAE& 0.3&  0.293&  0.239          \\
 SRCC &    0.92&  0.920&  0.943\\ \hline
 \multicolumn{4}{|c|}{BW-20K, 16 bar}\\ \hline
 MAE& 0.74&  0.73&  0.62          \\
 SRCC &    0.99&  0.99&  0.993\\ \hline
 \multicolumn{4}{|c|}{CoRE-2019, 0.15 bar}\\ \hline
 MAE& 0.54&  0.54&  0.49          \\
 SRCC &    0.84&  0.83&  0.85\\ \hline
 \multicolumn{4}{|c|}{CoRE-2019, 16 bar}\\ \hline
 MAE & 0.58 &  1.0 &  0.93           \\
 SRCC &    0.98 &  0.92 &  0.93\\ \hline
 \end{tabular} 
\end{center}
\caption{Table assessing the test performance of our enhanced crystal graph convolutional neural network (ECGCNN) for four adsorption data sets and two test-train splits.}
\label{tab:CoRE19}
\end{table}

\section{Discussion and Future Work}

In this paper we introduce a novel Enhanced Crystal Graph Convolutional Neural Network that rivals the performance of the state-of-the-art classical machine learning approaches in the literature~\cite{moosavi2020understanding}. In the referenced article, the authors construct a complex feature set of 165 chemical and geometric descriptors before deploying classical machine learning regression techniques to train their model. Our simpler model does not require the computation of a feature set and relies purely on known atomic properties and the information already encoded within the input file---the same information that is ingested by physics-based GCMC simulations for predicting adsorption properties.

This result is achieved despite the fact that our model does not fully account for gas molecule specificity. Future work needs to be done to incorporate the long-range inter-atomic potential between the $\textrm{CO}_2$ molecules and the framework atoms into our network. In particular, this would require the network to provide an accurate prediction of the final positions and orientations of the adsorbed gas molecules in the solid crystalline materials. The accomplishment of this task would represent a significant step beyond existing graph convolution models for crystalline materials which only account for the constituent atoms of the solid crystals and neglect any interaction with unbound gas or liquid molecules. The set of pairwise inter-atomic interactions is the last significant piece of physical information encoded within the simulation input file that is not utilised in our current model.

The limitations in the inclusion of gas molecule specificity in our ECGCNN model may explain the variations in its overall performance on the data sets presented in Table~\ref{tab:CoRE19}. In the benchmark paper~\cite{moosavi2020understanding}, the relative importance of the 165 chemical and geometric descriptors indicates that chemical specificity is more important for the CoRE-2019 data set relative to the BW-20K data set. 
%These chemical features are especially influential at lower as opposed to higher pressures. 
The observations could explain why our ECGCNN metrics in Table~\ref{tab:CoRE19} are better for the BW-20K data set over the CoRE-2019 data set.
%, and also why the network performance is superior at pressures of 0.15 bar as opposed to the 16 bar adsorption simulation results.      

Future work beyond refining the interaction between CO$_{2}$ molecules and the constituent atoms within the network, could include the calibration of the model to additional adsorption target properties. In particular, investigating the evolution of these properties within varying pressure or temperature ranges would allow us to extract engineering metrics that are vital for process engineering optimisation~\cite{farmahini2021performance}.

\section{Conclusions}
We have constructed an enhanced crystal graph convolution neural network (ECGCNN) to predict the carbon dioxide adsorption properties of MOF crystalline nanoporous materials. Our ECGCNN framework matches the best reported performance of classical machine learning methods, which require the costly computation of more than a hundred chemical and geometric material features. Future work could include greater gas molecule specificity within the network, as well as the training of the model to chemical process figures-of-merit for increased efficiency.
\section{Acknowledgements}
This material is based upon work supported in part by the
U.S. Department of Energy, Office of Science, Office of Advanced
Scientific Computing Research, under contract number
DE-AC05-00OR22725, and in part by the Laboratory Directed Research and
Development Program of Oak Ridge National Laboratory, managed by
UT-Battelle, LLC.

This research used resources of the Compute and Data Environment for
 Science (CADES) at the Oak Ridge National Laboratory, which is
 supported by the Office of Science of the U.S. Department of Energy
 under Contract No. DE-AC05-00OR22725.
 
This work was also supported by the Hartree National Centre for Digital Innovation, a collaboration between STFC and IBM.
\bibliography{aaai22}

\begin{thebibliography}{27}
\providecommand{\natexlab}[1]{#1}

\bibitem[{Boyd and Woo(2016)}]{boyd2016generalized}
Boyd, P.~G.; and Woo, T.~K. 2016.
\newblock A generalized method for constructing hypothetical nanoporous
  materials of any net topology from graph theory.
\newblock \emph{CrystEngComm}, 18(21): 3777--3792.

\bibitem[{Chen et~al.(2019)Chen, Ye, Zuo, Zheng, and Ong}]{chen2019graph}
Chen, C.; Ye, W.; Zuo, Y.; Zheng, C.; and Ong, S.~P. 2019.
\newblock Graph networks as a universal machine learning framework for
  molecules and crystals.
\newblock \emph{Chemistry of Materials}, 31(9): 3564--3572.

\bibitem[{Chen et~al.(2020)Chen, Wei, Huang, Ding, and Li}]{Chen2020gcnii}
Chen, M.; Wei, Z.; Huang, Z.; Ding, B.; and Li, Y. 2020.
\newblock Simple and Deep Graph Convolutional Networks.
\newblock In III, H.~D.; and Singh, A., eds., \emph{Proceedings of the 37th
  International Conference on Machine Learning}, volume 119 of
  \emph{Proceedings of Machine Learning Research}, 1725--1735. PMLR.

\bibitem[{Cheng, Zhang, and Dong(2021)}]{cheng2021geometric}
Cheng, J.; Zhang, C.; and Dong, L. 2021.
\newblock A geometric-information-enhanced crystal graph network for predicting
  properties of materials.
\newblock \emph{Communications Materials}, 2(1): 1--11.

\bibitem[{Chung et~al.(2019)Chung, Haldoupis, Bucior, Haranczyk, Lee, Zhang,
  Vogiatzis, Milisavljevic, Ling, Camp et~al.}]{chung2019advances}
Chung, Y.~G.; Haldoupis, E.; Bucior, B.~J.; Haranczyk, M.; Lee, S.; Zhang, H.;
  Vogiatzis, K.~D.; Milisavljevic, M.; Ling, S.; Camp, J.~S.; et~al. 2019.
\newblock Advances, updates, and analytics for the computation-ready,
  experimental metal--organic framework database: CoRE MOF 2019.
\newblock \emph{Journal of Chemical \& Engineering Data}, 64(12): 5985--5998.

\bibitem[{Chung and Lawson(2019)}]{chung2019persistence}
Chung, Y.-M.; and Lawson, A. 2019.
\newblock Persistence curves: A canonical framework for summarizing persistence
  diagrams.
\newblock \emph{arXiv preprint arXiv:1904.07768}.

\bibitem[{Coudert and Fuchs(2016)}]{coudert2016computational}
Coudert, F.-X.; and Fuchs, A.~H. 2016.
\newblock Computational characterization and prediction of metal--organic
  framework properties.
\newblock \emph{Coordination Chemistry Reviews}, 307: 211--236.

\bibitem[{Deeg et~al.(2020)Deeg, Damasceno~Borges, Ongari, Rampal, Talirz,
  Yakutovich, Huck, and Smit}]{deeg2020silico}
Deeg, K.~S.; Damasceno~Borges, D.; Ongari, D.; Rampal, N.; Talirz, L.;
  Yakutovich, A.~V.; Huck, J.~M.; and Smit, B. 2020.
\newblock In silico discovery of covalent organic frameworks for carbon
  capture.
\newblock \emph{ACS Applied Materials \& Interfaces}, 12(19): 21559--21568.

\bibitem[{Dureckova et~al.(2019)Dureckova, Krykunov, Aghaji, and
  Woo}]{dureckova2019robust}
Dureckova, H.; Krykunov, M.; Aghaji, M.~Z.; and Woo, T.~K. 2019.
\newblock Robust machine learning models for predicting high CO$_2$ working
  capacity and CO$_2$/H$_2$ selectivity of gas adsorption in metal organic
  frameworks for precombustion carbon capture.
\newblock \emph{The Journal of Physical Chemistry C}, 123(7): 4133--4139.

\bibitem[{Farmahini et~al.(2021)Farmahini, Krishnamurthy, Friedrich, Brandani,
  and Sarkisov}]{farmahini2021performance}
Farmahini, A.~H.; Krishnamurthy, S.; Friedrich, D.; Brandani, S.; and Sarkisov,
  L. 2021.
\newblock Performance-based screening of porous materials for carbon capture.
\newblock \emph{Chemical Reviews}, 121(17): 10666--10741.

\bibitem[{He et~al.(2016)He, Zhang, Ren, and Sun}]{he2016deep}
He, K.; Zhang, X.; Ren, S.; and Sun, J. 2016.
\newblock Deep residual learning for image recognition.
\newblock In \emph{Proceedings of the IEEE conference on computer vision and
  pattern recognition}, 770--778.

\bibitem[{Jablonka et~al.(2020)Jablonka, Ongari, Moosavi, and
  Smit}]{jablonka2020big}
Jablonka, K.~M.; Ongari, D.; Moosavi, S.~M.; and Smit, B. 2020.
\newblock Big-data science in porous materials: materials genomics and machine
  learning.
\newblock \emph{Chemical Reviews}, 120(16): 8066--8129.

\bibitem[{Kipf and Welling(2016)}]{kipf2016semi}
Kipf, T.~N.; and Welling, M. 2016.
\newblock Semi-supervised classification with graph convolutional networks.
\newblock \emph{arXiv preprint arXiv:1609.02907}.

\bibitem[{Krishnapriyan, Haranczyk, and
  Morozov(2020)}]{krishnapriyan2020topological}
Krishnapriyan, A.~S.; Haranczyk, M.; and Morozov, D. 2020.
\newblock Topological descriptors help predict guest adsorption in nanoporous
  materials.
\newblock \emph{The Journal of Physical Chemistry C}, 124(17): 9360--9368.

\bibitem[{Lee et~al.(2017)Lee, Barthel, D{\l}otko, Moosavi, Hess, and
  Smit}]{lee2017quantifying}
Lee, Y.; Barthel, S.~D.; D{\l}otko, P.; Moosavi, S.~M.; Hess, K.; and Smit, B.
  2017.
\newblock Quantifying similarity of pore-geometry in nanoporous materials.
\newblock \emph{Nature Communications}, 8(1): 1--8.

\bibitem[{Lee et~al.(2018)Lee, Barthel, D{\l}otko, Moosavi, Hess, and
  Smit}]{lee2018high}
Lee, Y.; Barthel, S.~D.; D{\l}otko, P.; Moosavi, S.~M.; Hess, K.; and Smit, B.
  2018.
\newblock High-throughput screening approach for nanoporous materials genome
  using topological data analysis: application to zeolites.
\newblock \emph{Journal of Chemical Theory and Computation}, 14(8): 4427--4437.

\bibitem[{Li et~al.(2021)Li, Mueller, Qian, Delgadillo~Perez, Abualshour,
  Thabet, and Ghanem}]{Li2021deepgcn}
Li, G.; Mueller, M.; Qian, G.; Delgadillo~Perez, I.~C.; Abualshour, A.; Thabet,
  A.~K.; and Ghanem, B. 2021.
\newblock DeepGCNs: Making GCNs Go as Deep as CNNs.
\newblock \emph{IEEE Transactions on Pattern Analysis and Machine
  Intelligence}, 1--1.

\bibitem[{Moosavi et~al.(2020)Moosavi, Nandy, Jablonka, Ongari, Janet, Boyd,
  Lee, Smit, and Kulik}]{moosavi2020understanding}
Moosavi, S.~M.; Nandy, A.; Jablonka, K.~M.; Ongari, D.; Janet, J.~P.; Boyd,
  P.~G.; Lee, Y.; Smit, B.; and Kulik, H.~J. 2020.
\newblock Understanding the diversity of the metal-organic framework ecosystem.
\newblock \emph{Nature Communications}, 11(1): 1--10.

\bibitem[{Park and Wolverton(2020)}]{park2020developing}
Park, C.~W.; and Wolverton, C. 2020.
\newblock Developing an improved crystal graph convolutional neural network
  framework for accelerated materials discovery.
\newblock \emph{Physical Review Materials}, 4(6): 063801.

\bibitem[{Rosen et~al.(2021)Rosen, Iyer, Ray, Yao, Aspuru-Guzik, Gagliardi,
  Notestein, and Snurr}]{rosen2021machine}
Rosen, A.~S.; Iyer, S.~M.; Ray, D.; Yao, Z.; Aspuru-Guzik, A.; Gagliardi, L.;
  Notestein, J.~M.; and Snurr, R.~Q. 2021.
\newblock Machine learning the quantum-chemical properties of metal--organic
  frameworks for accelerated materials discovery.
\newblock \emph{Matter}, 4(5): 1578--1597.

\bibitem[{Townsend et~al.(2020)Townsend, Micucci, Hymel, Maroulas, and
  Vogiatzis}]{townsend2020representation}
Townsend, J.; Micucci, C.~P.; Hymel, J.~H.; Maroulas, V.; and Vogiatzis, K.~D.
  2020.
\newblock Representation of molecular structures with persistent homology for
  machine learning applications in chemistry.
\newblock \emph{Nature Communications}, 11(1): 1--9.

\bibitem[{Unke and Meuwly(2019)}]{unke2019physnet}
Unke, O.~T.; and Meuwly, M. 2019.
\newblock PhysNet: a neural network for predicting energies, forces, dipole
  moments, and partial charges.
\newblock \emph{Journal of Chemical Theory and Computation}, 15(6): 3678--3693.

\bibitem[{Velickovic et~al.(2018)Velickovic, Cucurull, Casanova, Romero,
  Lio’, and Bengio}]{Velickovic2018GraphAN}
Velickovic, P.; Cucurull, G.; Casanova, A.; Romero, A.; Lio’, P.; and Bengio,
  Y. 2018.
\newblock Graph Attention Networks.
\newblock \emph{ArXiv}, abs/1710.10903.

\bibitem[{Wang et~al.(2020)Wang, Zhong, Bi, Yang, and
  Xu}]{wang2020accelerating}
Wang, R.; Zhong, Y.; Bi, L.; Yang, M.; and Xu, D. 2020.
\newblock Accelerating Discovery of Metal--Organic Frameworks for Methane
  Adsorption with Hierarchical Screening and Deep Learning.
\newblock \emph{ACS Applied Materials \& Interfaces}, 12(47): 52797--52807.

\bibitem[{Wilmer et~al.(2012)Wilmer, Leaf, Lee, Farha, Hauser, Hupp, and
  Snurr}]{wilmer2012large}
Wilmer, C.~E.; Leaf, M.; Lee, C.~Y.; Farha, O.~K.; Hauser, B.~G.; Hupp, J.~T.;
  and Snurr, R.~Q. 2012.
\newblock Large-scale screening of hypothetical metal--organic frameworks.
\newblock \emph{Nature Chemistry}, 4(2): 83--89.

\bibitem[{Xie and Grossman(2018)}]{xie2018crystal}
Xie, T.; and Grossman, J.~C. 2018.
\newblock Crystal graph convolutional neural networks for an accurate and
  interpretable prediction of material properties.
\newblock \emph{Physical Review Letters}, 120(14): 145301.

\bibitem[{Zhang et~al.(2019)Zhang, Cui, Zhang, Wu, and Lee}]{zhang2019machine}
Zhang, X.; Cui, J.; Zhang, K.; Wu, J.; and Lee, Y. 2019.
\newblock Machine learning prediction on properties of nanoporous materials
  utilizing pore geometry barcodes.
\newblock \emph{Journal of Chemical Information and Modeling}, 59(11):
  4636--4644.

\end{thebibliography}

\end{document}